\documentclass[8.5pt,twoside,twocolumn]{article}
\usepackage[super,sort&compress,comma]{natbib}
\usepackage{mhchem}
\usepackage{times,mathptmx}
\usepackage{sectsty}
\usepackage{balance}
\usepackage{xspace}
\usepackage{color}

\usepackage{graphicx} 
\usepackage{lastpage}
\usepackage[format=plain,justification=raggedright,singlelinecheck=false,font=small,labelfont=bf,labelsep=space]{caption}
\usepackage{fancyhdr}
\newcommand{\Or}{\ensuremath{\mathcal{O}}\xspace}
\newcommand{\jump}[1]{\big[\hspace{-0.7mm} \big[ #1 \big]
  \hspace{-0.7mm} \big]}
\newcommand{\mean}[1] {\big\{ \hspace{-0.7mm} \big\{ #1 \big\}
  \hspace{-0.7mm} \big\}}
\newcommand{\abs}[1]{\left\lvert#1\right\rvert}

\newcommand{\average}[1]{\left\langle#1\right\rangle}

\newcommand{\mc}[1]{\mathcal{#1}}
\newcommand{\DG}{\mathrm{DG}}
\newcommand{\eff}{\mathrm{eff}}

\begin{document}

\thispagestyle{plain} \fancypagestyle{plain}{
\renewcommand{\headrulewidth}{1pt}}
\renewcommand{\thefootnote}{\fnsymbol{footnote}}
\renewcommand\footnoterule{\vspace*{1pt}%
\hrule width 3.4in height 0.4pt \vspace*{5pt}}
\setcounter{secnumdepth}{5}

\makeatletter
\def\subsubsection{\@startsection{subsubsection}{3}{10pt}{-1.25ex plus -1ex minus -.1ex}{0ex plus 0ex}{\normalsize\bf}}
\def\paragraph{\@startsection{paragraph}{4}{10pt}{-1.25ex plus -1ex minus -.1ex}{0ex plus 0ex}{\normalsize\textit}}
\renewcommand\@biblabel[1]{#1}
\renewcommand\@makefntext[1]%
{\noindent\makebox[0pt][r]{\@thefnmark\,}#1} \makeatother
\renewcommand{\figurename}{\small{Fig.}~}
\sectionfont{\large}
\subsectionfont{\normalsize}

\fancyfoot{}
\fancyfoot[RO]{\footnotesize{\sffamily{1--\pageref{LastPage}
~\textbar  \hspace{2pt}\thepage}}}
\fancyfoot[LE]{\footnotesize{\sffamily{\thepage~\textbar\hspace{3.45cm}
1--\pageref{LastPage}}}} \fancyhead{}
\renewcommand{\headrulewidth}{1pt}
\renewcommand{\footrulewidth}{1pt}
\setlength{\arrayrulewidth}{1pt} \setlength{\columnsep}{6.5mm}
\setlength\bibsep{1pt}

\twocolumn[
  \begin{@twocolumnfalse}
\noindent\LARGE{\textbf{Edge reconstruction in armchair phosphorene
nanoribbons revealed by discontinuous Galerkin density functional
theory}} \vspace{0.6cm}

\noindent\large{\textbf{Wei Hu,$^{\ast}$\textit{$^{a}$} Lin
Lin,$^{\ast}$\textit{$^{ba}$} and Chao
Yang$^{\ast}$\textit{$^{a}$}}}\vspace{0.5cm}

\noindent\textit{\small{\textbf{Received Xth XXXXXXXXXX 20XX,
Accepted Xth XXXXXXXXX 20XX\newline First published on the web Xth
XXXXXXXXXX 200X}}}

\noindent \textbf{\small{DOI: 00.0000/00000000}} \vspace{0.6cm}

\noindent \normalsize{With the help of our recently developed
massively parallel DGDFT (Discontinuous Galerkin Density Functional
Theory) methodology, we perform large-scale Kohn-Sham density
functional theory calculations on phosphorene nanoribbons with
armchair edges (ACPNRs) containing a few thousands to ten thousand
atoms. The use of DGDFT allows us to systematically achieve
conventional plane wave basis set type of accuracy, but with a much
smaller number (about 15) of adaptive local basis (ALB) functions
per atom for this system. The relatively small number degrees of
freedom required to represent the Kohn-Sham Hamiltonian, together
with the use of the pole expansion the selected inversion (PEXSI)
technique that circumvents the need to diagonalize the Hamiltonian,
result in a highly efficient and scalable computational scheme for
analyzing the electronic structures of ACPNRs as well as its
dynamics. The total wall clock time for calculating the electronic
structures of large-scale ACPNRs containing 1080-10800 atoms is only
10-25 s per self-consistent field (SCF) iteration, with accuracy
fully comparable to that obtained from conventional planewave DFT
calculations. For the ACPNR system, we observe that the DGDFT
methodology can scale to 5,000-50,000 processors. We use DGDFT based
ab-initio molecular dynamics (AIMD) calculations to study the
thermodynamic stability of ACPNRs. Our calculations reveal that a 2
$\times$ 1 edge reconstruction appears in ACPNRs at room
temperature.}

\vspace{0.5cm} \end{@twocolumnfalse} ]

\footnotetext{\textit{$^{a}$Computational Research Division,
Lawrence Berkeley National Laboratory, Berkeley, CA 94720, USA}}
\footnotetext{\textit{$^{b}$Department of Mathematics, University of
California, Berkeley, CA 94720, USA}} \footnotetext{\textit{E-mail:
whu@lbl.gov (Wei Hu), linlin@math.berkeley.edu (Lin Lin),
cyang@lbl.gov (Chao Yang) }}

\section{Introduction}

Kohn-Sham density functional theory (DFT)\cite{PR_136_B864_1964_DFT,
PR_140_A1133_1965_DFT} is the most widely used methodology for
performing ab initio electronic structure calculations to study the
structural and electronic properties of molecules, solids and
nanomaterials. However, until recently, DFT calculations are limited
to small systems because they have a relatively high complexity
($\mathcal{O}(N^{2-3})$) with the system size $N$. As the system
size increases, the cost of traditional DFT calculations becomes
prohibitively expensive. Therefore, it is still challenging to use
DFT calculations to treat large-scale systems that may contain
thousand or tens of thousands of atoms. Although various linear
scaling $\mathcal{O}(N^{1})$ methods\cite{RMP_71_1085_1999_ON,
IRPC_29_665_2010_ON, RPP_75_036503_2010_ON} have been proposed for
improving the efficiency of DFT calculations, they rely on the
nearsightedness principle, which leads to exponentially localized
density matrices in real-space for systems with a finite energy gap
or at finite temperature.  On the other hand, most of the existing
linear scaling DFT codes, such as
SIESTA,\cite{JPCM_14_2745_2002_SIESTA}
CONQUEST,\cite{CPC_177_14_2007_CONQUEST}
OPENMX\cite{PRB_72_045121_2005_OPENMX} and
HONPAS,\cite{IJAC_2014_HONPAS} are based on the contracted and
localized basis sets in the real-space, such as Gaussian-type
orbitals or numerical atomic orbitals.\cite{IRPC_29_665_2010_ON}It
is relatively difficult to improve the accuracy of methods based on
such contracted basis functions in a systematic fashion compared to
methods based on conventional uniform basis sets, for example, the
planewave basis set,\cite{PRB_54_11169_1996_PlaneWave}. The
disadvantage of using uniform basis sets is the relatively large
number of basis functions required per atom.

Recently, we have developed a massively parallel DGDFT
(Discontinuous Galerkin Density Functional Theory) methodology for
performing efficient large-scale Kohn-Sham DFT calculations. The
methodology is based on the combination of the adaptive local basis
(ALB) set\cite{JCP_231_2140_2012_DGDFT}  and the pole expansion and
selected inversion (PEXSI) technique.\cite{LinLuYingEtAl2009,
JPCM_25_295501_2013_PEXSI, JPCM_26_305503_2014_PEXSI} The ALB
functions are localized in the real space and discontinuous in the
global domain. The continuous Kohn-Sham orbitals and density are
assembled from the discontinuous basis functions using the
discontinuous Galerkin (DG) method.\cite{SIAM_19_742_1982_DG,
ArnoldBrezziCockburnEtAl2002} Because it is rooted in a domain
decomposition approach that takes the chemical environment effects
into account, the ALB set constructed by the DGDFT methodology is
systematically improvable. It can achieve the same level of accuracy
obtained by conventional plane wave calculations with much fewer
number of basis functions. The sparse Hamiltonian matrix generated
from DGDFT can take advantage of the PEXSI method. The PEXSI method
overcomes the $\Or(N^3)$ scaling limit for solving Kohn-Sham DFT,
and scales at most as $\Or(N^2)$ even for metallic systems at room
temperature. In particular, the computational complexity of the
PEXSI method is only $\Or(N)$ for 1D systems, and is $\Or(N^{1.5})$
for 2D systems. This also makes the DGDFT methodology particularly
suitable for analyzing low-dimensional (1D and 2D) systems
regardless whether the system is a metal, a semiconductor or an
insulator.\cite{JPCM_25_295501_2013_PEXSI}

In this paper, we demonstrate the accuracy and efficiency
of DGDFT by using it to analyze the electronic structures and
thermodynamic stability of armchair phosphorene nanoribbons
(ACPNRs), which is an interesting 1D derivative of phosphorene with
some remarkable properties. We use DGDFT to perform both static
electronic structure calculations as well as ab initio molecular
dynamics (AIMD) calculations. Our AIMD calculations reveal that a $2
\times 1$ edge reconstruction appears in the edge unpassivated
ACPNRs at room temperature.

The paper is organized as follows. In section~\ref{sec:dgdft}, we
introduce our recently developed massively parallel DGDFT
methodology for efficient large-scale Kohn-Sham DFT based electronic
structure calculations. In section~\ref{sec:model}, we provide some
background on phosphorene nanoribbons that we examine. We report the
results obtained from applying DGDFT to ACPNRs in
section~\ref{sec:results}. We demonstrate that the DGDFT methodology
can achieve high accuracy with much fewer basis functions compared
to the conventional planewave discretized calculations. We also show
that DGDFT can handle large ACPNRs systems with thousand or even tens of
thousands of atoms. Furthermore, we show that the DGDFT methodology
is highly scalable on modern high performance computers because it
contains multiple levels of parallelization. Finally, we show that by
using DGDFT based ab-initio molecular dynamics (AIMD) calculations,
we are able to identify a 2 $\times$ 1 edge reconstruction in the
edge-unpassivated ACPNRs at room temperature. This observation
suggests that PNRs may modify their electronic structures over time,
hence are suitable phosphorene-based candidate materials for
nanoelectronics.

\section{DGDFT Methodology} \label{sec:dgdft}


In this section, we briefly present the mathematical foundation and
algorithmic ingredients of the DGDFT methodology. DGDFT constructs
adaptive local basis set (ALB) in the discontinuous Galerkin (DG)
framework.\cite{JCP_231_2140_2012_DGDFT} We explain why the
implementation of DGDFT is highly scalable on massively parallel
computers. Because the sparse Hamiltonian constructed by DGDFT can
take full advantage of the recently developed pole expansion and
selected inversion (PEXSI) method~\cite{LinLuYingEtAl2009,
JPCM_25_295501_2013_PEXSI, JPCM_26_305503_2014_PEXSI} to overcome
the $\Or(N^3)$ scaling of diagonalization methods, it can be used to
study the electronic structures and ab initio molecular dynamics
(AIMD) of large-scale atomistic systems.

\subsection{Adaptive local basis set in a discontinuous Galerkin
framework}

\begin{figure}[htbp]
\begin{center}
\includegraphics[width=0.4\textwidth]{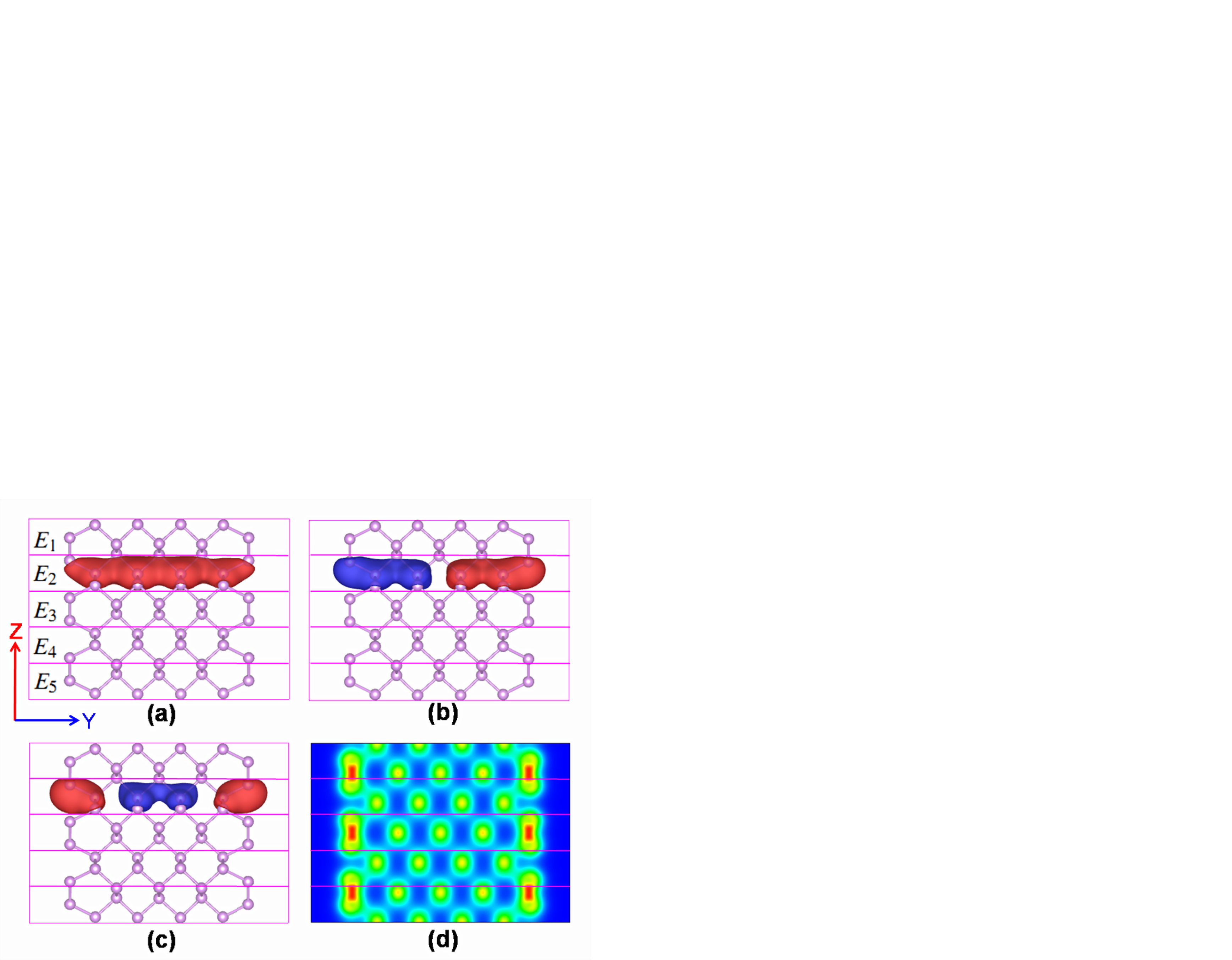}
\end{center}
\caption{(Color online) The isosurface of the first three ALB
functions, (a) $\phi$$_1$, (b) $\phi$$_2$, (c) $\phi$$_3$, belonging
to the second element and (d) the electron density $\rho$ across the
YZ plane in the global domain in the example of P$_{54}$. There are
5 elements and 160 ALB functions in each element in the P$_{54}$
system.} \label{fig:ALB}
\end{figure}

In our recent work,\cite{JCP_231_2140_2012_DGDFT} we have presented
a new way to discretize the Kohn-Sham Hamiltonian, called the
adaptive local basis functions (ALB). The basic idea of ALB is to
use eigenfunctions of the Kohn-Sham Hamiltonian defined on local domains to
construct basis functions. Compared to atom-centered basis functions
such as Gaussian type orbitals and numerical atomic orbitals, such
procedure encodes not only atomic structure but also environmental
effects into the basis functions.  In practice, we partition the
global computational domain into a number of subdomains (called
elements).  Then we define a buffer area for each element that
typically includes its nearest neighbor elements. We refer to the element
together with its buffer area as an extended element.  For instance,
Fig.~\ref{fig:ALB} shows an ACPNR with $54$ P atoms (P$_{54}$
system) partitioned along the Z-direction into $5$ elements. The
extended element associated with the second element $E_{2}$ contains
elements $E_{1},E_{2},E_{3}$, and the extended element
associated with the third element $E_{3}$ contains elements
$E_{2},E_{3},E_{4}$ and so on. We compute eigenfunctions for a local
Kohn-Sham problem in each extended element with periodic boundary
conditions using a local planewave basis set.  The artificial effect due
to the periodic boundary condition of the extended element is reduced by
restricting the point-wise values of eigenfunctions from the extended
element to the element, and the restricted eigenfunctions are mutually
orthogonalized on the element. We call such orthogonalized functions
adaptive local basis
functions. Note that the ALB functions can be computed at each step
of the self-consistent field (SCF) iteration through an efficient
iterative eigensolver using e.g. locally optimal block
preconditioned conjugate gradient (LOBPCG).\cite{Knyazev2001}

Since the elements are disjoint from each other, each ALB is
strictly zero outside its element, and is not continuous across the
boundaries of different elements.  Therefore,  we use the
discontinuous Galerkin (DG) method\cite{SIAM_19_742_1982_DG,
ArnoldBrezziCockburnEtAl2002} to construct a finite dimensional
Kohn-Sham Hamiltonian represented by these types of discontinuous
basis functions.  For instance, for periodic systems in a
norm-conserving
pseudopotential framework, the linearized DG energy functional at each step of
the self-consistent field (SCF) iteration becomes
\begin{equation}\label{eqn:DGvar}
  \begin{split}
    E_{\DG}(\{\psi_i\}) = &\frac{1}{2} \sum_{i=1}^N \average{\nabla
    \psi_i , \nabla \psi_i}_{\mc{T}}
    + \average{ V_{\eff}, \rho }_{\mc{T}}  \\
    &+ \sum_{I=1}^{N_A}\sum_{\ell=1}^{L_{I}} \gamma_{I,\ell} \sum_{i=1}^N
    \abs{\average{b_{I,\ell}(\cdot-R_{I}), \psi_i}_{\mc{T}}}^2 \\
    &- \sum_{i=1}^N
    \average{\mean{\nabla\psi_i}, \jump{\psi_i}}_{\mc{S}}
    + \alpha \sum_{i=1}^N \average{\jump{\psi_i},
    \jump{\psi_i}}_{\mc{S}}.
  \end{split}
\end{equation}
Here $\mc{T}$ is the collection of all elements (in
Fig.~\ref{fig:ALB} $\mc{T}=\left\{E_1, E_2, E_3, E_4, E_5 \right\}$,
with the collection of all its surfaces denoted by $\mc{S}$. The set
$\{\psi_{i}\}_{i=1}^{N}$ contains the $N$ occupied Kohn-Sham
orbitals represented as the linear combination of ALB functions. We
use $V_{\eff}$ to denote the effective one-body potential (including
local pseudopotential, Hartree potential and the
exchange-correlation potential) at each SCF iteration.  The terms
that contain $b_{I,\ell}$ and $\gamma_{I,\ell}$ correspond to the
nonlocal pseudopotential.  Here $\average{\cdot, \cdot}_{\mc{T}}$ is
the sum of the inner product on each element, and $\average{\cdot,
\cdot}_{\mc{S}}$ is the sum of the inner product on each surface.
$\mean{\cdot}$ and $\jump{\cdot}$ are the average and the jump
operators across surfaces due to the discontinuity of the basis
functions. We refer the readers to
Ref.\cite{JCP_231_2140_2012_DGDFT} for more detailed information.
What distinguishes the DG formulation from the standard Kohn-Sham
formulation of the DFT problem is the last two terms in
Eq.~\eqref{eqn:DGvar}, which comes from the integration by parts of
the Laplacian operator, and a penalty term to stabilize the numerical
evaluation of the energy, respectively. The DG method modifies the
Kohn-Sham energy
functional so that the kinetic energy functional is well defined
even with discontinuous basis functions.  The DG solution is also
fully consistent with the solution of standard Kohn-Sham equations
in the limit of a complete basis set, and the error can be measured
by a posteriori error estimators.\cite{KayeLinYang2015} The ALB
functions can achieve high accuracy (less than $1$ meV per atom) in
the total energy calculation with a very small number (4$-$40) of
basis functions per atom, compared to fully converged planewave
calculations.

Using a 1D ACPNR (P$_{54}$) as an example, we show the isosurfaces
of the first three ALB functions in the second element in
Fig.~\ref{fig:ALB}(a)-(c). The global computational domain is
partitioned along the Z-direction into $5$ elements. Each ALB
function shown is strictly localized inside the second element and
is therefore discontinuous across the boundary of elements. On the
other hand, each ALB function is delocalized across a few atoms
inside the element since they are obtained from eigenfunctions of
local Kohn-Sham Hamiltonian. Although the basis functions are
discontinuous, the electron density is well-defined and is very
close to be a continuous function in the global domain
(Fig.~\ref{fig:ALB}(d)) once the local contributions are assembled.
It should be noted that
all ALB functions are by construction mutually orthogonal. Thus the
corresponding overlap matrix is an identity matrix. Hence, this
formulation avoids solving a generalized eigenvalue problem that has
a potentially ill-conditioned overlap matrix.

\subsection{Two levels parallelization strategy}

The DGDFT framework naturally allows two levels of
parallelization. For each element, the computation of eigenfunctions
for the local Kohn-Sham Hamiltonian can be parallelized similar to how a
regular Kohn-Sham DFT solver with planewave basis sets is parallelized.
This type of fine-grained parallelization is called intra-element
parallelization. On top of this, the
computation of eigenfunctions for different elements, together with
the construction of the DG Hamiltonian can be naturally
parallelized at a coarse grain level. This is called inter-element
parallelization. We optimized the data communication structure so that different levels
of parallelization can be seamlessly and efficiently performed. We
will demonstrate the details of the parallelization strategy on
massively parallel computers in a separate publication in
preparation.\cite{DGDFT_Parallelization_2015}

In the intra-element parallelization, the wavefunction and eigenfunctions of
each extended element are distributed among different processors. The
number of eigenfunctions to be computed for a single element is
usually on the order of $100$, and intra-element parallelization can
scale to several hundred processors.  The level of concurrency that
can be achieved in the inter-element parallelization is determined by
the number of elements. In the DGDFT method, each element usually
takes around $10$ atoms, and for a system containing $1000$ atoms
there should be around $100$ elements. As a result, the two-level
parallelization strategy can readily scale to $10,000$ processors.
For the largest ACPNRs system studied in this work, the number of
processors used is 50,000 processors.

\subsection{Pole expansion and selected inversion method}

Once the DG Hamiltonian is constructed, one can solve a standard eigenvalue
problem to obtain physical quantities such as
electron density, total energy and atomic forces. This can be done
by treating the DG Hamiltonian matrix as a dense matrix and by
solving the eigenvalue problem via standard parallel linear algebra software
packages for dense matrices, e.g. ScaLAPACK\cite{ScaLAPACK}
(referred to as the ``DIAG'' method). The computational cost of the
DIAG method scales as $\Or(N^3)$. This parallel scalability of ScaLAPACK
diagonalization subroutine is limited to a few thousands of processors.
When more than $10,000$ processors are available, DIAG can become
the computational bottleneck because it cannot take advantage of that
many processors even though other part of the DGDFT calculation become
less time consuming.

The recently developed pole expansion~pole expansion and selected
inversion (PEXSI) method\cite{LinLuYingEtAl2009,
JPCM_25_295501_2013_PEXSI, JPCM_26_305503_2014_PEXSI} avoids the
diagonalization procedure completely. It evaluates physical
quantities such as electron density, energy, atomic force without
calculating any eigenvalue or eigenfunction, and reduces the
computational complexity to at most $\Or(N^2)$ without sacrificing
accuracy even for metallic systems. In particular, the computational
complexity of the PEXSI method is only $\mathcal{O}(N)$ for 1D
systems (such as ACPNRs studied here), and is $\Or(N^{1.5})$ for 2D
systems. These are much more favorable compared with the
$\mathcal{O}(N^{3})$ complexity of the DIAG method. Therefore, the
PEXSI method is particularly suitable to study the electronic
structure of larges scale  low-dimensional (1D and 2D) systems. The
PEXSI method is also highly scalable to more than $10,000$
processors, as recently demonstrated in the massively parallel
SIESTA-PEXSI method\cite{JPCM_26_305503_2014_PEXSI,
JCP_141_214704_2014_GNFs} based on numerical atomic orbitals.
Therefore the combined DGDFT-PEXSI method can scale beyond $10,000$
processors and solves electronic structure problem with more than
$10,000$ atoms.

\section{Theoretical model of ACPNRs} \label{sec:model}

Phosphorene, a new two dimensional (2D) elemental
monolayer,\cite{ACSNano_8_4033_2014, NatureCommun_5_4475_2014,
NatureNanotech_9_372_2014, JPCL_5_1289_2014} has received
considerable amount of interest recently after it has been
experimentally isolated through mechanical exfoliation from bulk
black phosphorus. Phosphorene exhibits some remarkable electronic
properties superior to graphene, a well known elemental
sp$^2$-hybridized carbon monolayer.\cite{Scinece_306_666_2004,
NatureMater_6_183_2007, RMP_81_109_2009} For example, phosphorene is
a direct semiconductor with a high hole
mobility.\cite{ACSNano_8_4033_2014} It has the drain current
modulation up to 10$^5$ in nanoelectronics.\cite{NatureCommun_5_4475_2014} These
remarkable properties have already been used for wide applications
in field effect transistors\cite{NatureNanotech_9_372_2014} and
thin-film solar cells.\cite{JPCL_5_1289_2014} Furthermore, up to
now, phosphorene is the only stable elemental 2D material which can
be mechanically exfoliated in experiments\cite{ACSNano_8_4033_2014}
besides graphene. Therefore, it can potentially be used as an
alternative to graphene\cite{ScienceNews_185_13_2014} in the future
and lead to faster semiconductor electronics.


By cutting 2D phosphorene into finite-sized 1D phosphorene
nanoribbons (PNRs), a bandgap engineering technique often used for
graphene\cite{PRL_99_216802_2007, Nature_6_652_2007,
NatureMater_6_770_2007} to get graphene nanoribbons
(GNRs),\cite{NanoLett_6_2748_2006, PRL_97_216803_2006,
PRL_99_186801_2007} one obtains a new type of material that has been
subject to many theoretical and experimental
studies.\cite{PRB_90_085424_2014, JPCC_118_14051_2014,
JPCL_5_2675_2014, NanoLett_14_4607_2014} The stability and
electronic properties of PNRs depend sensitively on the ribbon width
and how it is cut from the 2D phosphorene, which can result in
either armchair or zigzag shaped edges.\cite{JPCC_118_14051_2014}
Unlike GNRs,\cite{NanoLett_6_2748_2006, PRL_97_216803_2006,
PRL_99_186801_2007} hydrogen-passivated PNRs with armchair and
zigzag edges are all semiconductors with direct band
gaps.\cite{JPCC_118_14051_2014} For edge-unpassivated PNRs armchair
edged PNRs (ACPNRs) are all semiconducting, but zigzag edged PNRs
(ZZPNRs) all exhibit metallic characteristics. Furthermore, it has
been found that edge-unpassivated ZZPNRs exhibit instability at the
edge boundary that may easily induce edge reconstruction and
disorder. Using density functional theory (DFT) calculations,
Ramasubramaniam et al.\cite{PRB_90_085424_2014} have shown that a 2
$\times$ 1 edge reconstruction appears in the edge-unpassivated
ZZPNRs. The reconstruction induces different stability and
electronic structures of ZZPNRs. Edge disorder is also observed by
Guo et al.\cite{JPCC_118_14051_2014} in the edge-unpassivated ZZPNRs
with ab-initio molecular dynamics calculations.

However, the edge-unpassivated ACPNRs seem to be thermodynamically
stable at the edge boundary,\cite{JPCC_118_14051_2014} and up to
now, no edge reconstruction or disorder has been predicted
theoretically in the edge-unpassivated ACPNRs. In the present work,
we focus on the edge-unpassivated ACPNRs because the
hydrogen-passivated PNRs been theoretically proved to be very
thermodynamically stable and the edge reconstruction has been
observed in the edge-unpassivated ZZPNRs.\cite{PRB_90_085424_2014}

\begin{figure}[htbp]
\begin{center}
\includegraphics[width=0.4\textwidth]{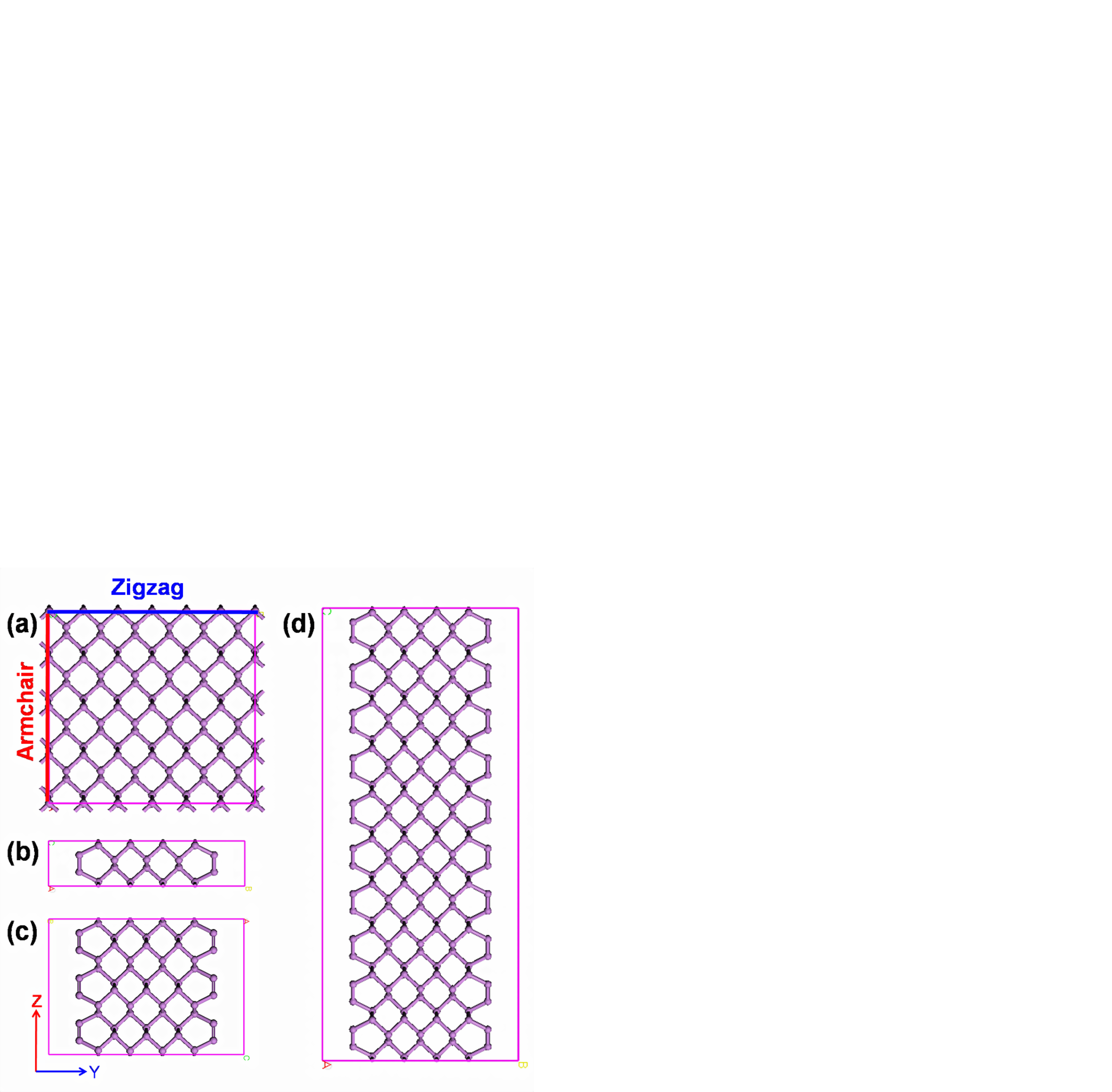}
\end{center}
\caption{(Color online) Geometric structures of (a) 2D phosphorene
in the 1 $\times$ 6 $\times$ 4 supercell and different 1D ACPNRs with a width
$N$ = 4 in the (b) unit cell (P$_{18}$), (c) 1 $\times$ 1 $\times$ 3 (P$_{54}$)
and (d) 1 $\times$ 1 $\times$ 10 (P$_{180}$) supercells. The violet balls
denote phosphorus atoms. Two types of edges, armchair and zigzag,
are highlighted in the insert.} \label{fig:Structure}
\end{figure}

Fig.~\ref{fig:Structure} shows the atomic configuration of a 2D
phosphorene monolayer in a 1 $\times$ 6 $\times$ 4 supercell and some examples
of 1D ACPNRs with a width $N$ = 4 in the unit cell (P$_{18}$), 1
$\times$ 1 $\times$ 3 (P$_{54}$) and 1 $\times$ 1 $\times$ 10 (P$_{180}$) supercells.
Other ACPNRs in very large supercells involving thousand or tens of
thousands of atoms, such as the 1 $\times$ 1 $\times$ 120 (P$_{2160}$), 1
$\times$ 1 $\times$ 240 (P$_{4320}$) and 1 $\times$ 1 $\times$ 600 (P$_{10800}$)
supercells, which we adopt in this work, are not shown here. The
vacuum space in the X and Y directions is about 10 {\AA} to separate
the interactions between neighboring slabs in ACPNRs.

\section{Results and Discussion} \label{sec:results}

In this section, we present computational results obtained
by applying DGDFT to ACPNRs of different sizes. We demonstrate
the accuracy of the calculation and parallel efficiency of
DGDFT. We also report a $2\times 1$ edge reconstruction
observed in a AIMD study performed to assess the thermodynamic
stability of ACPNRs.

We use the conventional plane wave software package
ABINIT\cite{CPC_180_2580_2009_ABINIT} as a reference to check the
accuracy for our DGDFT calculations. The same exchange-correlation
functionals, including the local density approximation of Goedecker,
Teter, Hutter (LDA-Teter93)\cite{PRB_54_1703_1996_LDA} and
generalized gradient approximation of Perdew, Burke, and Ernzerhof
(GGA-PBE),\cite{PRL_77_3865_1996_PBE} and the
Hartwigsen-Goedecker-Hutter (HGH) norm-conserving
pseudopotential\cite{PRB_58_3641_1998_HGH} are adopted in both
ABINIT and DGDFT software packages. All calculations are
performed on the Edison system available at the National Energy
Research Scientific Computing (NERSC) center.

\subsection{Computational accuracy}

We first check the accuracy of total energy and atomic force of the
DGDFT software package by using P$_{54}$ shown in Figure~\ref{fig:Structure}(c)
as an example. To simplify our discussion, we define the
total energy error per atom $\Delta$E (Hartree/atom) and maximum
atomic force error $\Delta$F (Hartree/Bohr) as
\[
\Delta{E}=(E^{\text{DGDFT}}-E^{\text{ABINIT}})/N
\] and
\[\Delta{F}=\max_I|F_I^{\text{DGDFT}}-F_I^{\text{ABINIT}}|\]
respectively, where $N$ and $I$ correspond to the total number of
atoms and an atom index, $E^{\text{DGDFT}}$ and
$E^{\text{ABINIT}}$ represent the total energy computed
by DGDFT and ABINIT respectively, and $F_I^{\text{DGDFT}}$ and
$F_I^{\text{ABINIT}}$ represent the Hellmann-Feynman force on the
$I$-th
phosphorus atom in P$_{54}$ computed by DGDFT
and ABINIT, respectively. We find that neglecting the Pulay force in
the atomic force leads to moderate deviation in the conserved energy in
the AIMD simulation.
The ABINIT results are obtained by setting
the energy cutoff to 200 Hartree for the wavefunction to ensure full
convergence.
The kinetic energy cutoff (denoted by Ecut) in the
DGDFT method is used to define the grid size for computing the ALBs as
is in standard Kohn-Sham DFT calculations using planewave basis sets.
Ecut is also directly related to the Legendre-Gauss-Lobatto (LGL) integration
grid defined on each element and used to perform numerical integration as
needed to construct the DG Hamiltonian matrix.

\begin{table}
\caption{
The accuracy of DGDFT in terms of the total energy error per atom
$\Delta$$E$ (Hartree/atom) and the maximum atomic force error
$\Delta$$F$ (Hartree/Bohr) in the DIAG and PEXSI methods with
different energy cutoff Ecut (Hartree) of wavefunction and number of
ALB functions per atom, compared with converged ABINIT calculations.
\#ALB means the number of ALB functions per atom.
} \label{Accuracy}
\begin{tabular}{cccccc} \\ \hline \hline
\multicolumn{2}{c}{DGDFT P$_{54}$}  & \multicolumn{2}{c}{DIAG}  & \multicolumn{2}{c}{PEXSI} \ \\
Ecut &  \#ALB  & $\Delta$$E$ & $\Delta$$F$ & $\Delta$$E$ & $\Delta$$F$  \ \\
\hline
10   &  28   & 1.94E-02  &  4.81E-02 & 1.94E-02  &  4.81E-02  \ \\
20   &  28   & 6.49E-04  &  5.12E-03 & 5.39E-04  &  1.67E-02  \ \\
40   &  10   & 1.28E-03  &  1.52E-02 & 1.21E-03  &  4.19E-03  \ \\
40   &  12   & 5.54E-04  &  2.17E-03 & 6.45E-04  &  2.17E-03  \ \\
40   &  15   & 1.87E-04  &  9.54E-04 & 1.16E-04  &  9.57E-04  \ \\
40   &  19   & 7.00E-05  &  4.00E-04 & 7.12E-05  &  4.13E-04  \ \\
40   &  28   & 9.64E-06  &  2.90E-04 & 4.21E-05  &  2.84E-04  \ \\
100  &  28   & 8.25E-06  &  1.24E-04 & 2.90E-05  &  1.31E-04  \ \\
200  &  28   & 6.62E-06  &  9.43E-05 & 3.66E-05  &  9.09E-05  \ \\
\hline \hline
\end{tabular}
\end{table}

Table~\ref{Accuracy} shows that the total energy and atomic forces
produced by the DGDFT method are highly accurate compared to the
ABINIT results.
In particular, the total energy error $\Delta E$
can be as small as $6.6 \times 10^{-6}$ Hartree/atom if the DIAG
method is used to compute the charge density and $3.7 \times
10^{-5}$ Hartree/atom if the  PEXSI method is used to compute the
charge density respectively. The maximum error of the atomic force can
be as small as $9.4 \times 10^{-5}$ Hartree/Bohr when DIAG is used and $9.1
\times 10^{-5}$ Hartree/Bohr when PEXSI is used. These results are
obtained when only a relatively small number (28) of ALB functions per atom are
used to construct the global DG Hamiltonian. The energy cutoff for
the local wavefunctions use to represent the ALB functions is set to 200
Hartree in this case. Note that the accuracy of total energy and
atomic force in DGDFT depends on both the energy cutoff for local
wavefunctions defined on an extended element and the number of ALB
functions.  We can see from Table~\ref{Accuracy} that the accuracy
in energy and forces both improve as either the energy cutoff or the
number of ALB functions increases. This clearly demonstrates that the ALB set
produced by the DGDFT methodology is systematically improvable.

When the PEXSI method\cite{LinLuYingEtAl2009,
JPCM_25_295501_2013_PEXSI, JPCM_26_305503_2014_PEXSI} is used to
compute the charge density, the accuracy of the computation is
determined by the number of poles used in the pole
expansion.\cite{JPCM_25_295501_2013_PEXSI} We examined the effect of
the number of poles on the accuracy of total energy and atomic force
in DGDFT, and found that sufficiently high accuracy (comparable to
that achieved by using the DIAG method to compute the charge
density) can be achieved when the number of poles is set to 50.

In our parallel efficiency tests and AIMD simulations, we use a
energy cutoff of 40 Hartree for wavefunction and 15 ALB functions
per atom to achieve a good compromise between accuracy and
computational efficiency.  For this particular choice of energy
cutoff and number of ALB functions, we are able to keep the total
energy error under 1 $\times$ 10$^{-4}$ Hartree/atom and atomic
force error under 1 $\times$ 10$^{-3}$ Hartree/Bohr for large-scale
ACPNRs.

\subsection{Parallel efficiency}

\begin{figure}[htbp]
\begin{center}
\includegraphics[width=0.4\textwidth]{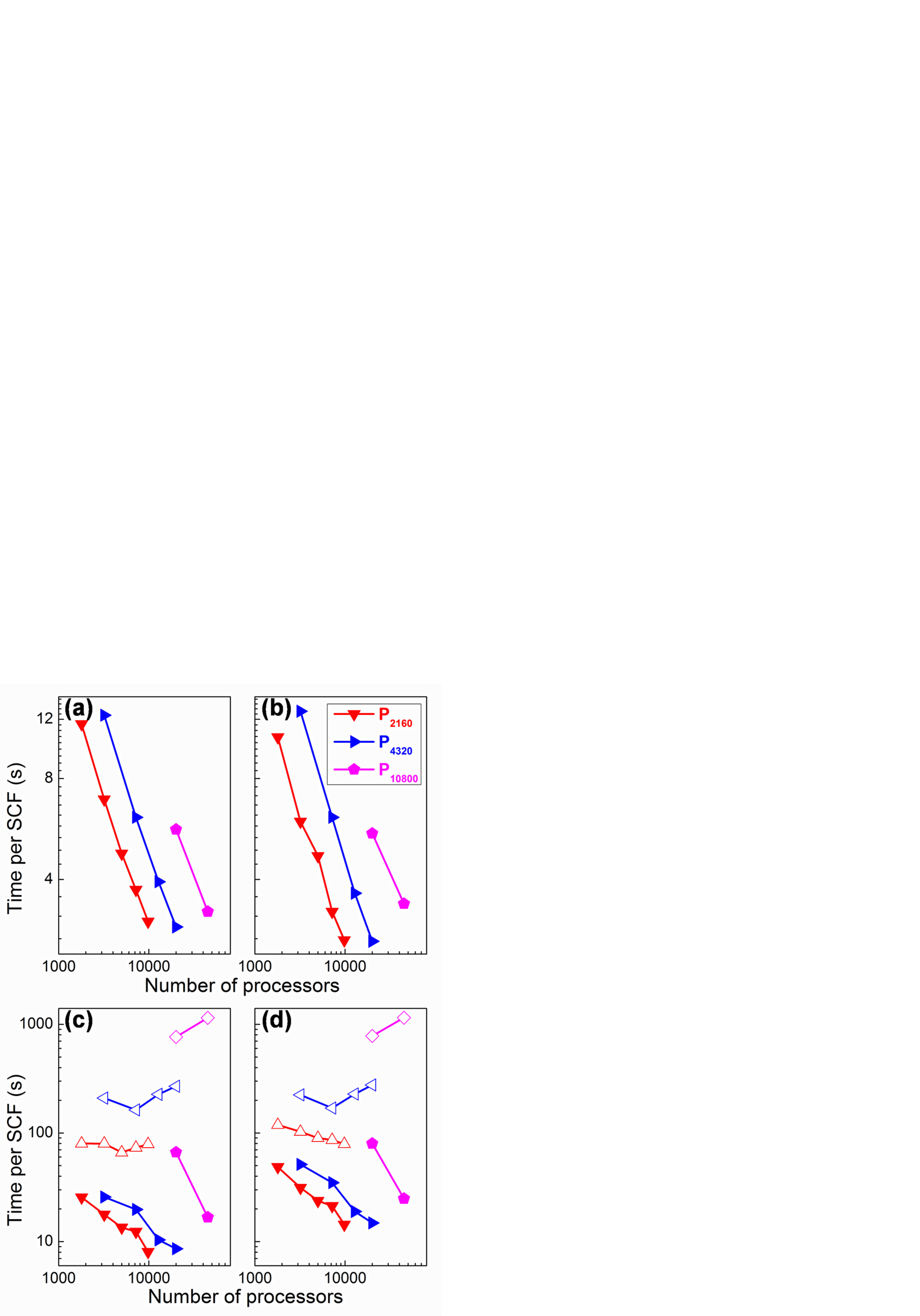}
\end{center}
\caption{(Color online)
The change of wall clock time with respect to the number of
processors used for the computation for three ACPNR systems of
different sizes ($P_{2160}$, $P_{4320}$ and $P_{10800}$). (a) Strong
scaling of the generation of ALB functions step, (b) strong scaling
of the DG Hamiltonian matrix construction step, (c) strong scaling
the evaluation of the approximate charge density, energy and forces
from the constructed DG Hamiltonian matrix, (d) strong scaling of
the overall computation. The reported wall clock time is for one SCF
iteration. The timing and scaling shown in (c) and (d) depend on
whether DIAG (hollow markers) or PEXSI (solid markers) is used to
evaluate physical quantities such as charge density, energy and
forces. } \label{fig:Parallelization}
\end{figure}

In the DGDFT method, each SCF iteration performs the following three
main steps of computation: (a) the generation of ALB functions, (b) the
construction of DG Hamiltonian matrix via ALB functions and (c)
the evaluation of the approximate charge density, energy and atomic forces
by either diagonalizing
the DG Hamiltonian (DIAG) or by using the PEXSI technique. Note that there are
some additional steps such as the computation of energy, charge mixing
or potential mixing, and intermediate data communication etc. The cost
of these steps is included in the total wall clock time in
Fig.~\ref{fig:Parallelization} (d).

Fig.~\ref{fig:Parallelization} shows the strong parallel scaling of
these three individual steps of computation, as well as the
overall computation, for three large scale ACPNRs
(P$_{2160}$, P$_{4320}$ and P$_{10800}$)
in terms of the wall clock time per SCF step.

The wall clock time of the first two steps are independent of
whether PEXSI or DIAG is used to evaluate electron density,
energy and forces.  Fig.~\ref{fig:Parallelization}(a) and
(b) show that they both scale nearly perfectly with respect
to the number of processors used in the computation for
all test problems we used. Furthermore,
The total wall clock time required to perform each one
these steps is reduced to a few seconds even for
P$_{10800}$ when more than 10,000 processors are used in the
computation.

Fig.~\ref{fig:Parallelization}(c) and (d) show that
the evaluation of the approximate charge density using the DG
Hamiltonian matrix dominates the total wall clock time per SCF iteration
in the DGDFT methodology. For large-scale ACPNRs, the PEXSI method can
effectively reduce the wall clock time compared to the DIAG method in the
DGDFT methodology. Furthermore, using the DIAG method with
ScaLAPACK,\cite{ScaLAPACK} appears to limit the strong parallel
scalability for large-scale ACPNRs to at most 5,000 processors on the Edison. Increasing the number of processors beyond
that can lead to an increase in wall clock time. In contrast,
the PEXSI method exhibits highly scalable performance. It can make
efficient use of more than 20,000-50,000 processors on Edison for $P_{10800}$.
It should be noted that the total wall clock time required for
performing large-scale ACPNRs containing thousands or tens of thousands
of atoms is only about 10-25 seconds per SCF iteration.

\subsection{AIMD simulation}

Ab-initio molecular dynamics (AIMD) simulation capability has been
implemented in the DGDFT method.\cite{DGDFT_MD_2015} We use DGDFT
AIMD simulation to study the thermodynamic stability of ACPNRs.
Using P$_{180}$ as an example, we perform an AIMD simulation to
obtain a 2.5 picosecond (ps) trajectory of ACPNR dynamics with a time step of 2.0
femtosecond (fs) under canonical ensemble with the temperature fixed
at 300 K controlled by a single level Nose-Hoover
thermostat.\cite{JCP_81_511_1984_Nose, PRA_31_1695_1985_Hoover} The
mass of the Nose-Hoover thermostat is chosen to be $85000$ au. We
use the GGA-PBE\cite{PRL_77_3865_1996_PBE} exchange-correlation
functional for this particular simulation.

\begin{figure}[htbp]
\begin{center}
\includegraphics[width=0.4\textwidth]{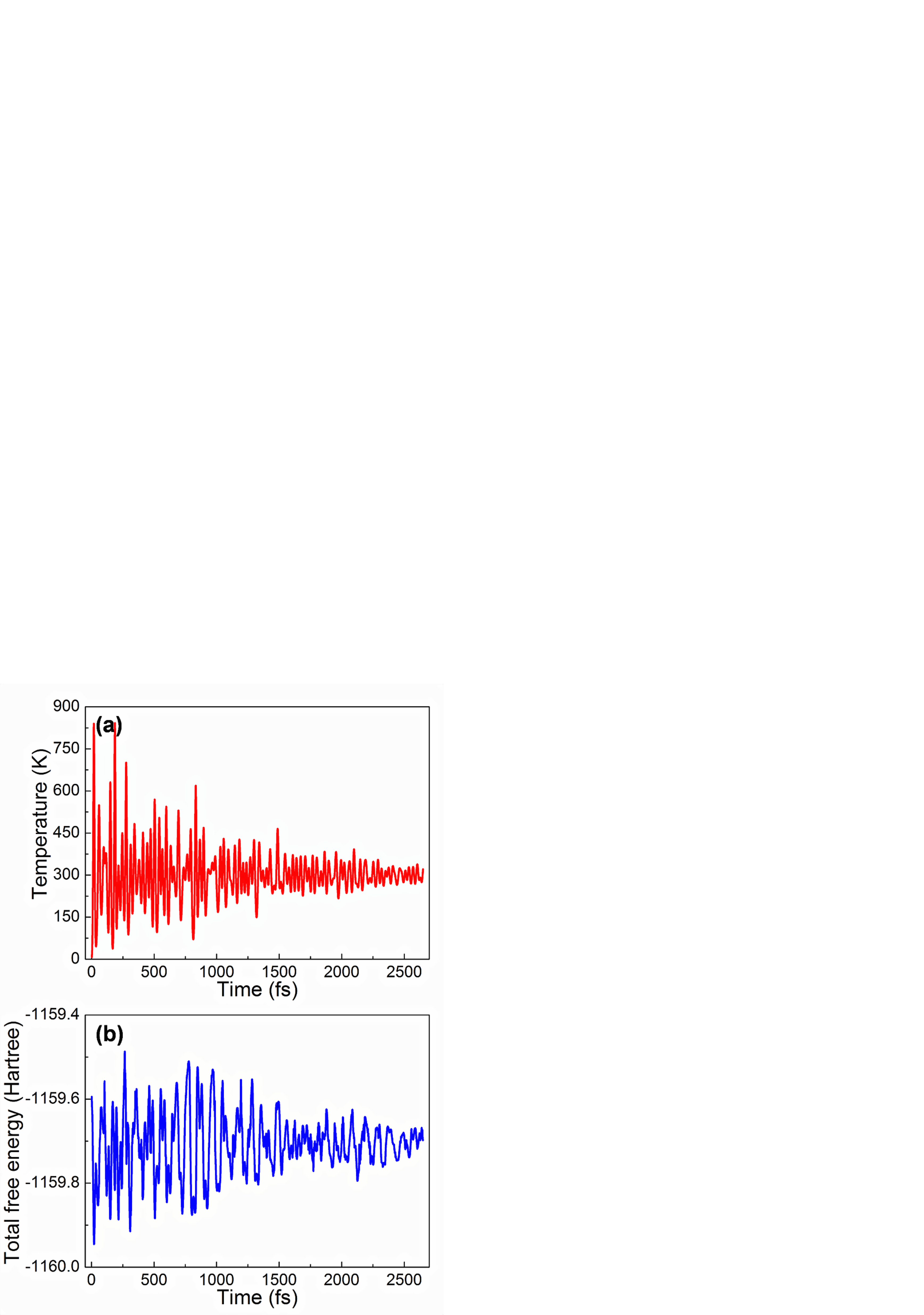}
\end{center}
\caption{(Color online) (a) kinetic temperature and (b) total free
energy along the AIMD trajectory for the ACPNR (P$_{180}$). The
simulation is performed for 2.5 ps at 300 K.
} \label{fig:MD}
\end{figure}

In Fig.~\ref{fig:MD}, we plot the temperature (computed by $3/2 N
k_B T = E_K$ where $E_{K}$ is the kinetic energy) and total free
energy of P$_{180}$ along the simulated trajectory. The temperature
of the system reaches around $300$ K after $1.5$ ps. Although DGDFT
only uses the Hellmann-Feynman force, we have observed that the drift
of the conserved Hamiltonian in the Nose-Hoover thermostat is
relatively small at $2.6\times 10^{-4}$ Hartree per atom per ps.

\begin{figure}[htbp]
\begin{center}
\includegraphics[width=0.4\textwidth]{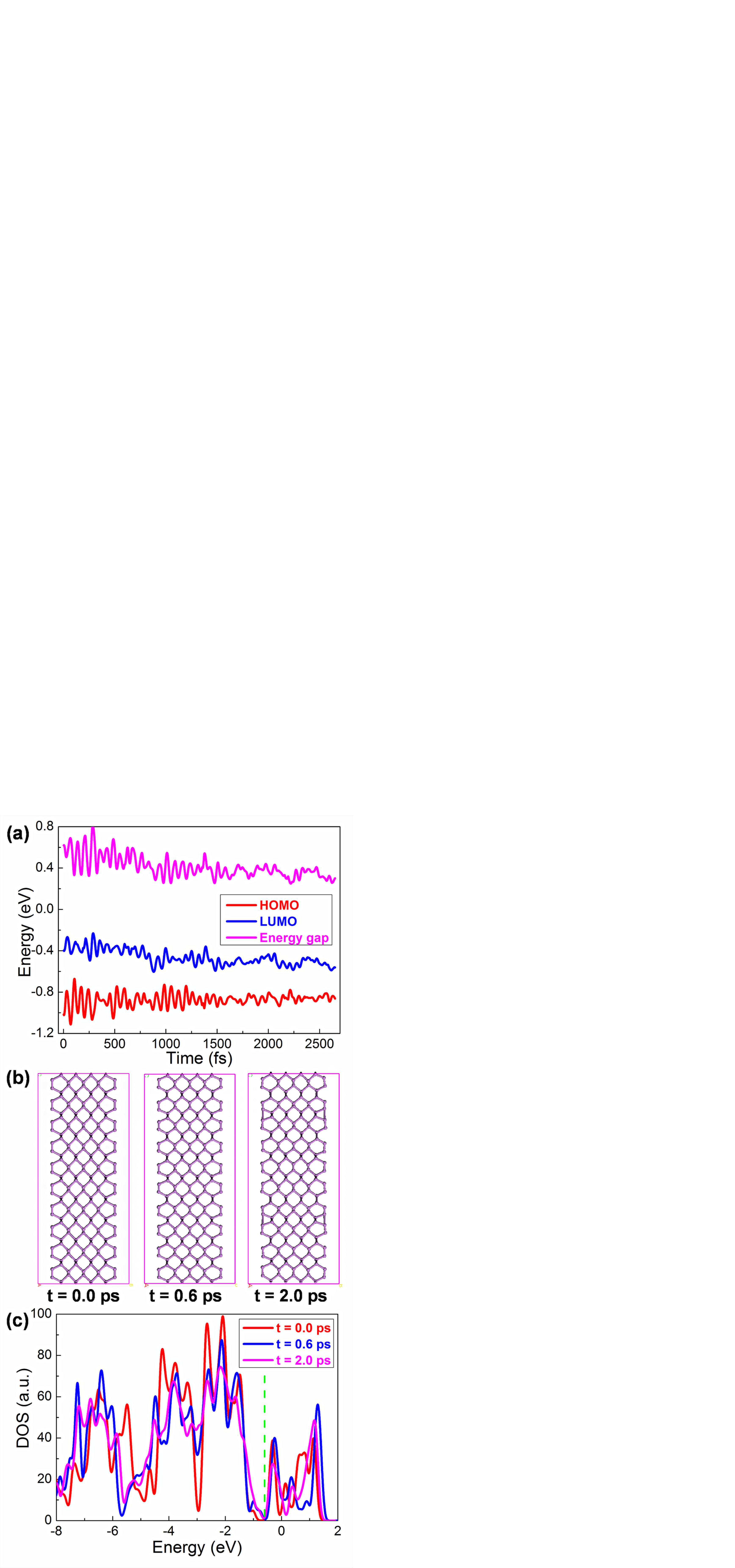}
\end{center}
\caption{(Color online)
(a) HOMO, LUMO and
energy gap for the ACPNR (P$_{180}$) along the AIMD trajectory. (b)
Geometric structures and (c) density of states (DOS) at three
snapshots of the simulation at $t$ = 0.0, 0.6 and 2.0 ps. The Fermi level
is marked by the green dotted line.
} \label{fig:DOS}
\end{figure}

We examine the electronic structures of P$_{180}$ during 2.5 ps at
300 K as shown in Fig.~\ref{fig:DOS}. Geometric structures and
density of states (DOS) of three AIMD snapshots at $t$ = 0.0, 0.6
and 2.0 ps are plotted in Fig.~\ref{fig:DOS}(b) and (c). In the
initial configuration (t = 0.0 ps), the geometry of ACPNR is
optimized first by using a gradient descent method with the
Barzilai-Borwein line search
technique\cite{SpringerUS_96_235_2005_BB} implemented in DGDFT.
After t = 0.6 ps, the ACPNR exhibits some local deformations due to
the thermal perturbation introduced by the temperature. After t =
2.0 ps, 2 $\times$ 1 edge reconstruction can be observed. We find
that the electronic structures of ACPNRs are also affected by
thermal perturbation and edge reconstruction. Fig.~\ref{fig:DOS} (a)
indicates that the highest occupied molecular orbital (HOMO) energy
is shifted by around $-0.3$ eV, and the lowest unoccupied molecular
orbital (LUMO) energy is shifted by around $-0.2$ eV along the MD
trajectory. The HOMO-LUMO  energy gaps of P$_{180}$ are calculated
to be 0.63, 0.44 and 0.38 eV at $t$ = 0.0, 0.6 and 2.0 ps,
respectively, showing that the shift of the energy level is more
pronounced for the HOMO than the LUMO as shown in Fig.~\ref{fig:DOS}
(c). Therefore, the edge-unpassivated ACPNRs are also
thermodynamically unstable just like the edge-unpassivated
ZZPNRs.\cite{PRB_90_085424_2014, JPCC_118_14051_2014} This behavior
is quite different from that of edge-unpassivated graphene
nanoribbons.\cite{NanoLett_6_2748_2006, PRL_97_216803_2006,
PRL_99_186801_2007} The reconstruction of edges in PNRs can modify
their electronic\cite{PRB_90_085424_2014, JPCC_118_14051_2014} and
transport\cite{JPCL_5_2675_2014} properties and make them potential
candidate materials for phosphorene-based electronic devices, such as
field effect transistors.\cite{NatureNanotech_9_372_2014}


\section{Conclusions}

In summary, we developed a massively parallel DGDFT (Discontinuous
Galerkin Density Functional Theory) methodology for efficient
large-scale Kohn-Sham density functional theory (DFT) calculations
based on the combination of the adaptive local basis (ALB) set and
the pole expansion and selected inversion (PEXSI) technique. The
DGDFT methodology can achieve a high basis set accuracy comparable
to that provided by conventional plane wave calculations but with a
small number of ALB basis functions per atom for large-scale
electronic structure calculations that involve thousand or tens of
thousands of atoms. Furthermore, the DGDFT methodology is highly
scalable based on two levels of parallelization (intra- and
inter-element parallelization), which can make efficient use of more
than 50,000 processors on high performance machines for the systems
studied here. Using ab-initio molecular dynamics calculations on
armchair phosphorene nanoribbons (ACPNRs), we find that a 2 $\times$
1 edge reconstruction appears in ACPNRs at room temperature to
modify their electronic structures for phosphorene-based
nanoelectronics in the future.

\section{Acknowledgments}

This work is partially supported by the Scientific Discovery through
Advanced Computing (SciDAC) Program funded by U.S. Department of
Energy, Office of Science, Advanced Scientific Computing Research
and Basic Energy Sciences (W. H., L. L. and C. Y.), and by the
Center for Applied Mathematics for Energy Research Applications
(CAMERA), which is a partnership between Basic Energy Sciences and
Advanced Scientific Computing Research at the U.S Department of
Energy (L. L. and C. Y.). We thank the National Energy Research
Scientific Computing (NERSC) center for the computational resources.

\footnotesize{
\bibliography{rsc} 
\bibliographystyle{rsc} 
}

\end{document}